\documentclass[aps,pra,showpacs,reprint,superscriptaddress,floatfix]{revtex4-1}
\usepackage{graphicx}
\usepackage{longtable}
\usepackage{dcolumn}
\usepackage{multirow}
\usepackage{amssymb, amsmath}
\usepackage[colorlinks=true, citecolor=blue,linkcolor=blue]{hyperref}
\begin{document}
\title{\emph{Ab initio} calculations of permanent dipole moments and dipole polarizabilities of alkaline-earth monofluorides}
\author{Renu Bala}
\email{rbala@ph.iitr.ac.in}
\author{H. S. Nataraj}
\email{hnrajfph@iitr.ac.in}
\affiliation{ Department of Physics, Indian Institute of Technology Roorkee,
Roorkee - 247667, India}
\author{Malaya K. Nayak}
\email{ mknayak@barc.gov.in, mk.nayak72@gmail.com}
\affiliation{Theoretical Chemistry Section, Chemistry Group, Bhabha Atomic Research Centre, Trombay Mumbai 400085, India}
\begin{abstract}
The ground\,-\,state permanent dipole moments (PDMs) and molecular dipole polarizabilities (DPs) of open-shell alkaline-earth monofluorides, and atomic DPs of alkaline\,-\,earth- and fluorine atoms are reported at the Kramers\,-\,restricted configuration interaction level of theory limited to single and double excitations (KRCISD), using the finite\,-\,field approach. Sufficiently large basis sets such as quadrupole\,-\,zeta (QZ) and augmented\,-\,QZ basis sets together with the generalized active space technique is employed to carry out the field dependent energy calculations at the KRCISD level. The PDMs and the components of DPs are extracted from the linear- and quadratic fit of energies against perturbative electric field, respectively. Accuracy of the present calculations for the electronic properties is examined by comparison with the measurements and calculations where ever  available.
\end{abstract}
\maketitle
%
\section{Introduction}
%
%
The alkaline-earth monofluorides (AEMFs) are currently in the forefront of several important theoretical as well as experimental research works not only in physics but also in chemistry. The AEMFs have attracted theorists because of their simple electronic structure: open-shell with their ground-state being $^2\Sigma^+$, and the experimentalists because of the ease with which they could be laser-cooled. The latter is evident from the multitude of experiments that have been carried out to realize cooling and trapping of MgF~\cite{Yin}, CaF~\cite{Tarbutt, Zhelyazkova} SrF~\cite{Barry, Barry1, Shuman, Truppe}, and BaF~\cite{Bu, Altunta} molecules. Some of the heavier members of AEMFs such as BaF~\cite{Nayak} and RaF~\cite{Kudashov, Isaev} have been studied chiefly for fundamental symmetry violating effects which give rise to nuclear anapole moment. Such effects bear vital implications for the physics beyond the Standard Model (SM) of particle physics. The chemical reaction between the SrF molecules at ultracold temperatures has been investigated by Meyer \emph{et al.}~\cite{Meyer} that can be used for the study of dipolar degenerate quantum gases. 

The accurate theoretical predictions of the  molecular electronic properties such as permanent dipole moments (PDMs) and dipole polarizabilities (DPs) are crucial in guiding the ongoing and future experiments with cold and ultracold AEMF molecules. It is worth mentioning that the knowledge of PDMs of molecules is helpful for the study of long\,-\,range dipole\,-\,dipole interactions, electric dipole moment of an electron (eEDM), for qubits in quantum computations, and for several quantum phases in ultracold gases~\cite{Menotti, DeMille, Meyer1, Trefzger, Goral}. Similarly, the knowledge of electric DPs is one of the requisites for manipulating and for controlling the dynamics of molecules in laser fields ~\cite{Friedrich, Friedrich1, Friedrich2, Friedrich3, Deiglmayr}.

A few theoretical studies of AEMF molecules that are being investigated in this work have already been reported in the literature. To mention: a detailed structural study of BaF molecule has been performed by Tohme and Korek~\cite{Tohme}; the spectroscopy and evaluation of properties such as dipole moments of SrF molecules for the ground- and low\,-\,lying excited states has been performed by Jardali~\emph{et al.}~\cite{Jardali}; the electronic structure calculations for the ground- and several excited-states of XF (X\,=\,Be, Mg, Ca) molecules have been performed by Kork~\emph{et al.}~\cite{Kork}. Further, the XF (X\,=\,Be, Mg, Ca, Sr, Ba) molecules together with mercury halides, and PbF have been studied by Abe~\emph{et al.}~\cite{Abe} recently for the PDMs, and for effective electric fields ($E_{eff}$) using relativistic finite\,-\,field coupled\,-\,cluster method with single and double excitations (FFCCSD). The theoretical results for average and anisotropic polarizability ($\bar{\alpha}$ and $\gamma$) components of BeF and MgF at the complete active space self\,-\,consistent field (CASSCF) level, and model-based predictions for AEMFs are available in Refs.~\cite{Fowler, Davis}. The polarizability components of SrF calculated using CCSD method with partial triples [CCSD(T)] has been reported in Refs.~\cite{Meyer, Kosicki} while, only the \textit{z}\,-\,component of polarizability for BaF has been calculated using multi\,-\,reference configuration interaction (MRCI) method in Ref.~\cite{Tohme}. Nevertheless, there are no experimental results available in the literature, known to our knowledge, for the DPs of these molecules. 

In this work, we have studied the whole series of group\,-\,II monofluorides systematically and uniformly using the same quality of basis sets, and employing the same level of correlation method. In addition to the PDMs ($\mu_0$), and components of molecular DPs ($\alpha_{\parallel}$ and $\alpha_{\bot}$) of AEMFs, we have also calculated the atomic polarizabilities ($\alpha_A$) of alkaline-earth- and fluorine atoms using generalized active space (GAS) technique at the CI level of theory. The accuracy of our results is analyzed by comparing them with the available results in the literature.

The paper is organized in four sections. The theory and the method employed to calculate the PDMs, and polarizabilities is discussed in Section~\ref{section-2}, followed by the results and discussions of these properties in Section~\ref{section-3}, and at last, the summary of this work in Section~\ref{section-4}.
\section{Theory and Method of Calculations}
\label{section-2}
In the presence of an external homogeneous electric field of strength $\varepsilon$, the total energy $E(\varepsilon)$ of a molecule can be written as a Taylor expansion:
\begin{eqnarray}{}
\displaystyle
E(\varepsilon)\,=\,E_0\,+\,\mu_0\, \varepsilon \,+\,\frac{1}{2}\,\alpha\,\varepsilon^2\,+\,.\,.\,.\,,
\end{eqnarray}
where,
\begin{eqnarray}{\label{PDM}}
\displaystyle
\mu_0\,=\,-\left(\frac{dE}{d\varepsilon}\right)_0,\; \mathrm{and} \;\;
\alpha\,=\,-\left(\frac{d^2E}{d\varepsilon^2}\right)_0.
\end{eqnarray}
Here, $\mu_0$ is the permanent dipole moment, and $\alpha$ is the dipole polarizability. The subscript $0$ on the field derivatives of the total energy indicate that these are evaluated at $\varepsilon \rightarrow 0$.

By considering $z$\,-\,axis as the internuclear axis of the molecule, we obtain two components of polarizability: the  component  ($\alpha_{\parallel}\equiv\alpha_{zz}$) parallel to the axis, and the  component ($\alpha_{\bot}\equiv\alpha_{xx}\equiv\alpha_{yy}$) perpendicular to the direction of the internuclear axis. The average ($\bar{\alpha}$), and the anisotropic polarizabilities ($\gamma$) are calculated in terms of these components as,

\begin{eqnarray}{\label{pol1}}
\displaystyle
\bar{\alpha} = (\alpha{_\parallel }+2\alpha{_\bot})/3\quad,\,\, \mathrm{and} \quad \gamma = \alpha{_\parallel}-\alpha{_\bot}.
\end{eqnarray}

In order to compute the foregoing electric properties, the relativistic field\,-\,dependent energies are calculated using Dirac\,-\,Fock (DF) and Kramers\,-\,restricted configuration interaction method with single and double excitations (KRCISD) for both alkaline-earth- and fluorine atoms, as well as for AEMF molecules. After generating the reference state using DF Hamiltonian, CISD calculations employing the GAS technique are performed using the KRCI module of DIRAC15 software suite~\cite{DIRAC}. The Gaussian charge distribution for the nuclei is used in these calculations. Further, we have used the uncontracted correlation\,-\,consistent polarized valence quadruple zeta\,(cc\,-\,pVQZ)~\cite{Dunning} basis sets for Be\,($12$s\,$6$p\,$3$d\,$2$f\,$1$g), F\,($12$s\,$6$p\,$3$d\,$2$f\,$1$g), and Mg\,($16$s\,$12$p\,$3$d\,$2$f\,$1$g), and Dyall basis sets of similar quality (dyall.v4z)~\cite{Dyall_basis} for Ca\,($30$s\,$20$p\,$6$d\,$5$f\,$3$g), Sr\,($33$s\,$25$p\,$15$d\,$4$f\,$3$g), Ba\,($35$s\,$30$p\,$19$d\,$4$f\,$3$g), and Ra\,($37$s\,$34$p\,$23$d\,$15$f\,$3$g). These basis sets are quite large, specially when used in uncontracted form, as it can be seen from the explicit functions shown in parenthesis. Such computationally expensive large basis sets are considered in this work for obtaining reliable results. The energies of the systems (both atoms and molecules) studied here are calculated using perturbative electric field in the range; ($2\,-\,5$)\,$\times\,10^{-4}$ E$_h$/ea$_0$. These calculations are repeated for computing different components along different axes. Using the first\,- and the second\,-\,order polynomial fits to the data of  $E(\varepsilon)$ against $\varepsilon$, we obtain PDMs and DPs, respectively.  

As the properties studied in this work are the valence properties, we have adopted a frozen-core approximation to perform the correlation calculations for atoms as well as for molecules. The atomic polarizability calculations are carried out with $10$ outermost electrons {\itshape viz.} $n$s,\,$n$p, and ($n\,+\,1$)s orbitals where $n$ being $2$ for Mg, $3$ for Ca, $4$ for Sr, $5$ for Ba, and $6$ for Ra atom. However, for Be and F atoms, all electrons are kept active in the post\,-\,DF calculations to correlate more or less the same number of electrons. Further, a virtual cutoff energy of $10\,E_h$ is set uniformly for all atoms and molecules in order to truncate the orbitals having higher energies so that the computations can be made manageable.

\begin{table}[ht]
\begin{ruledtabular}
\begin{center}
\caption{\label{table-I} Generalized active space model for the CI wavefunctions of atoms with $10E_h$ virtual cutoff energy.}
\begin{tabular}{c|ccccccc}
Subspace & \multicolumn{7}{c}{Number of orbitals}\\
\cline{2-8}
 & Be & Mg & Ca & Sr & Ba & Ra & F\\
\hline
Frozen core & 0 & 1 & 5 & 14 & 23 & 39 & 0 \\
GAS1 & 1 & 4 & 4 & 4 & 4 & 4 & 1 \\
GAS2 & 1 & 1 & 1 & 1 & 1 & 1 & 4 \\
GAS3 & 56 & 59 & 114 & 107 & 110 & 108 & 25\\
GAS3$^*$ & 81 & 84 & 139 & 132 & 135 & 133 & 50\\
  \end{tabular}
\begin{flushleft}
$^*$Number of virtual orbitals in case of augmented basis sets.
\end{flushleft}
\end{center}
\end{ruledtabular}
\end{table}
The atomic orbitals  are partitioned into four subspaces: frozen core, GAS$1$, GAS$2$ and GAS$3$ is shown in Table~\ref{table-I}. The first subspace is the frozen core and it is excluded in correlation calculations. For Be atom, filled $1$s and $2$s orbitals form GAS$1$ and GAS$2$ subspace, respectively. However, for other alkaline-earth atoms GAS$1$ and GAS$2$ subspaces contain active [$n$s,\,$n$p], and [$(n+1)$s] orbitals, respectively. On the other hand, for the open shell F atom, $1$s orbital form GAS$1$ and [$2$s,\,$2$p] orbitals form GAS$2$ subspace.  All electrons available in GAS$1$ and GAS$2$ subspaces are allowed to excite to the virtual orbitals contained in GAS$3$ subspace. In case of augmented basis sets, GAS3$^*$ replaces GAS3.

\begin{table}[ht]
\begin{ruledtabular}
\begin{center}
\caption{\label{table-II} Generalized active space model for the CI wavefunctions of AEMFs with $10E_h$ virtual cutoff energy.}
\begin{tabular}{c|cccccc}
Subspace& \multicolumn{6}{c}{Number of orbitals}\\
\cline{2-7}
        & BeF & MgF & CaF & SrF & BaF & RaF \\
\hline
Frozen core &  2 & 6 & 7 & 15 & 24 & 40\\
GAS1 & 4 & 4 & 7 & 8 & 8 & 8  \\
GAS2 & 1 & 1 & 1 & 1 & 1  & 1\\
GAS3 & 80 & 84 & 139 & 132 & 134 & 133 \\
GAS3$^*$ & 128 & 132 & 188 & 180 & 183 & 183\\
  \end{tabular}
\begin{flushleft}
$^*$Number of virtual orbitals in case of augmented basis sets.
\end{flushleft}
\end{center}
\end{ruledtabular}
\end{table}
For molecular property calculations, the DF orbitals having energy less than -$2\,E_h$ are considered as frozen core. The alkaline-earth atom is chosen as the coordinate origin of the corresponding diatomic molecule. In the GAS technique, active DF orbitals are divided into three subspaces: paired (GAS1), unpaired (GAS2), and virtual orbitals (GAS3), as shown in Table~\ref{table-II}.

The values of equilibrium bond lengths used in the present work are: $1.359$ {\AA} for BeF~\cite{Kork}, $1.778$ {\AA} for MgF~\cite{Kork}, $2.015$ {\AA} for CaF~\cite{Kork}, $2.124$ {\AA} for SrF~\cite{Jardali}, $2.162$ {\AA} for BaF~\cite{Tohme}, and $2.244$ {\AA} for RaF~\cite{Isaev}. Atomic units for distance ($1$\,a$_0$\,$=\,0.52917721$\,{\AA}), electric field ($1$\,E$_h$/ea$_0$\,$=\,5.142\times10^{11}$\,V/m), dipole moment ($1$\,ea$_0$\,$=\,2.54174691$\,D), and dipole polarizability ($1$\,e$^2$a$_0^2$/E$_h$\,$=\,0.14818474$\,{\AA}$^3$) are used throughout the paper.
\begin{table*}[ht!]
\begin{ruledtabular}
\begin{center}
\caption{\label{table-III} Results of DPs for the ground-state of alkaline-earth- and fluorine atoms compared with the available results in the literature.}
\begin{tabular}{llllllll}
Method & Be & Mg & Ca & Sr & Ba & Ra & F\\
\hline
DF/QZ & 45.5 & 81.0 & 183.0 & 233.0 & 324.0 & 297.0 & 2.5 \\
DF/aug\,-\,QZ & 45.5 & 80.5 & 182.5 & 232.0 & 323.5 & 297.0 & 3.5\\
KRCISD/QZ & 37.0 & 73.0 & 157.0& 194.0 & 269.5 & 247.5 & 3.0 \\
KRCISD/aug\,-\,QZ & 38.5& 73.0& 157.5 & 196.5& 269.0& 248.5& 3.5\\
  \hline
 Expt. & $-$ & 70.89$^a$, 59\,(16)$^b$ & 169\,(17)$^c$ & 186\,(15)$^d$ & 268\,(22)$^c$ & $-$ & $-$\\
 PRCC~\cite{Chattopadhyay} &$-$ & 70.76 & 160.77 & 190.82 & 274.68 & 242.42 & $-$\\
 CCSD~\cite{Sahoo} & 37.80\,(47) & 73.41\,(2.32) & 154.58\,(5.42) & 199.71\,(7.28)& 268.19\,(8.74) & $-$ & $-$\\
 CI\,+\,MBPT$2$~\cite{Porsev} & 37.76\,(22) & 71.3\,(7) & 157.1\,(1.3) & 197.2\,(2) & 273.5\,(2.0) &$-$ &$-$\\
 DF~\cite{Lim} &$-$ &$-$ & 182.79 & 232.66 & 323.82 & 299.59 & $-$\\
 CCSD(T)$_{recommended}$~\cite{Lim} & $-$&$-$ & 157.9 & 199 & 273.5 & 246.2 & $-$\\
 CCSD(T) & $-$ & $-$ & $-$ & $-$  & 272.7$^e$ & 242.8$^e$ & 3.70$^f$\\
 CASPT2$^*$ & 37.2$^g$ & 70.9$^g$ & 163$^g$ & 210$^g$ & 312$^g$ &283$^g$ &3.76$^h$\\
 CASSCF~\cite{Nelin} & $-$ & $-$ & $-$ & $-$ & $-$ &$-$ & 3.68\\
 MR\,-\,CCI$^{\dag}$~\cite{Nelin} & $-$ & $-$ & $-$ & $-$ & $-$ &$-$ & 3.52\\
  \end{tabular}
\begin{flushleft}
 $^a$Reference~\cite{Peter}  $^b$Reference~\cite{Ma}\\
  $^c$Reference~\cite{Miller} $^d$Reference~\cite{Miller1} \\ 
 $^e$Reference~\cite{Borschevsky} $^f$Reference~\cite{Das}\\
  $^g$Reference~\cite{Roos} $^h$Reference~\cite{Medved}\\
 $^*$Complete\,-\,active\,-\,space second\,-\,order perturbation theory.\\
 $^{\dag}$Externally contracted multireference configuration interaction.
\end{flushleft}
\end{center}
\end{ruledtabular}
\end{table*}
%
\section{\label{section-3}Results and Discussion}
\begin{table}[ht]
\begin{ruledtabular}
\begin{center}
\caption{\label{table-IV} PDM (in Debye) for the ground-state of AEMFs at equilibrium bond length with QZ basis sets, compared with the results in the literature.}
\begin{tabular}{llll}
%
Molecule & Method & $\mu_0$ & Ref.\\
  \hline
 BeF & DF &  1.246& This work\\
     & KRCISD & 1.116& This work\\
     & FFCCSD & 1.15 & \cite{Abe}\\
     & LCCSD$^a$&  1.10 & \cite{Prasannaa}\\
     & FD\,-\,HF & 1.2727 & \cite{Kobus}\\
     & CISD & 1.131 & \cite{Langhoff}\\
     & CPF & 1.086 & \cite{Langhoff}\\   
     & MP2 & 1.197 & \cite{Buckingham}\\
     \hline
 MgF & DF & 3.207& This work\\
      & KRCISD  & 3.124& This work\\
      & FFCCSD & 3.13 & \cite{Abe}\\
      & LCCSD$^a$  & 3.07 & \cite{Prasannaa}\\
      & FD\,-\,HF  & 3.1005 & \cite{Kobus}\\
      & CISD & 3.048 & \cite{Langhoff}\\
      & CPF  & 3.077 & \cite{Langhoff}\\
      & MP2 & 3.186 & \cite {Buckingham}\\
      & IM $^b$ & 3.64 & \cite{Torring}\\            
      & EPM$^c$ & 3.5 & \cite{Mestdagh}\\
      \hline
 CaF & DF &  2.894& This work\\
     & KRCISD &  3.181 & This work\\
     & FFCCSD & 3.19 & \cite{Abe}\\     
     & LCCSD$^a$   & 3.16 & \cite{Prasannaa}\\
     & FD\,-\,HF & 2.6450 & \cite{Kobus}\\
     & CISD & 2.590 & \cite{Langhoff}\\ 
     & CPF & 3.060 & \cite{Langhoff}\\
     & MP2 & 3.190 & \cite{Buckingham}\\ 
     & IM$^b$ & 3.34  & \cite{Torring}\\
     & LFA$^d$ & 3.00& \cite{Rice}\\
     & EPM$^c$ & 3.2 & \cite{Mestdagh}\\     
     & MRD\,-\,CI$^e$ & 3.01 & \cite{Bundgen}\\
     & LFA$^c$ & 3.36 & \cite{Allouche}\\
     & Exp. & 3.07(7) & \cite{Childs}\\
 \hline
 SrF & DF &  2.930 & This work\\
     & KRCISD  & 3.395 & This work\\     
     & FFCCSD & 3.62 & \cite{Abe}\\ 
     & CCSD(T) & 3.4564 & \cite{Kosicki}\\
     & LCCSD$^a$  & 3.60 & \cite{Prasannaa}\\
     & Z-vector & 3.4504 & \cite{Sasmal}\\
     & FD-HF & 2.5759 & \cite{Kobus}\\
     & CISD & 2.523 & \cite{Langhoff}\\
     & CPF & 3.199 & \cite{Langhoff}\\ 
     & IM$^b$ & 3.67 & \cite{Torring}\\ 
     & EPM$^c$ & 3.6 & \cite{Mestdagh}\\
     & LFA$^d$ & 3.79 & \cite{Allouche}\\ 
     & Exp. &  3.4676(10) & \cite{Ernst}\\                    
  \end{tabular}
\begin{flushleft}
\end{flushleft}
\end{center}
\end{ruledtabular}
\end{table}     
     \setcounter{table}{3}
 \begin{table}[t]
 \begin{ruledtabular}
 \begin{center}
 \caption{ \label{table-IV} Continued...}
 \begin{tabular}{llll} 
 Molecule & Method &  $\mu_0$ & Ref.\\   
  \hline
  BaF & DF  &2.122 & This work \\
      & KRCISD  & 2.706& This work\\ 
      & MRCISD & 2.958 & \cite{Tohme}\\ 
      & FFCCSD & 3.41 & \cite{Abe}\\       
      & LCCSD$^a$  & 3.40 & \cite{Prasannaa}\\
      & IM$^b$ & 3.44 &\cite{Torring}\\
      & EPM$^c$ & 3.4 & \cite{Mestdagh}\\      
      & LFA$^d$ & 3.91 & \cite{Allouche}\\
      & Exp. &  3.170(3) & \cite{Ernst1}\\
 \hline
 RaF & DF  &3.045 & This work\\
     & KRCISD  & 3.621& This work\\
  \end{tabular}
\begin{flushleft}
$^a$Linearized coupled\,-\,cluster method with single and double excitations\\ $^b$Ionic model\\ $^c$Electrostatic polarization model\\ $^d$Ligand\,-\,field approach\\ $^e$Multi\,-\,reference single\,- and double excitation configuration\,-\,interaction\\
 \end{flushleft}
\end{center}
\end{ruledtabular}
\end{table}
\begin{table}[ht]
\begin{ruledtabular}
\begin{center}
\caption{\label{table-V} PDM (in Debye) for the ground-state of AEMFs at equilibrium bond length with aug-QZ basis sets.}
\begin{tabular}{lll}
Molecule & Method & $\mu_0$  \\
\hline
BeF & DF & 1.246  \\
    & KRCISD & 1.143  \\
MgF & DF & 3.235 \\
    & KRCISD & 3.177 \\
CaF & DF & 2.893 \\
    & KRCISD & 3.221 \\
SrF & DF &2.963 \\
    & KRCISD & 3.465 \\
BaF & DF & 2.122  \\  
RaF & DF & 3.050\\  
  \end{tabular}
\begin{flushleft}  
\end{flushleft}
\end{center}
\end{ruledtabular}
\end{table}
\begin{table*}[t]
\begin{ruledtabular}
\begin{center}
\caption{\label{table-VI} Results of DPs for the ground-state of AEMFs at equilibrium bond length with QZ basis sets. Comparison with the available results in the literature.}
\begin{tabular}{lllllll}
Molecule & Method &  $\alpha$$_{\parallel}$ & $\alpha$$_{\bot }$ & $\bar{\alpha}$ & $\gamma$ & Ref.\\
  \hline
 BeF & DF &  18.0 & 29.0 & 25.33 & -11.0 & This work\\
     & KRCISD &  19.0 & 29.5 & 26.0 & -10.5 & This work\\
     & CASSCF &  $-$ & $-$ & 27.30 & -10.45 & \cite{Fowler}\\
     & T-Rittner & $-$ & $-$ & 1.69 & $-$ & \cite{Davis}\\
     & TEK$^a$ & $-$ & $-$ &1.79& -0.19 &\cite{Davis}\\
     & D-shell\,(-1)$^b$& $-$ & $-$ &1.92& 0.19 &\cite{Davis}\\
     & D-shell\,(q)$^c$ & $-$ & $-$ &2.04& 1.34 &\cite{Davis}\\
     \hline
 MgF  & DF &  34.5 & 57.5 & 49.83 & -23.0 & This work\\
      & KRCISD &  37.5 & 58.0 & 51.16 & -20.5 & This work\\
      & CASSCF &  $-$ & $-$ & 53.34 & -22.79 & \cite{Fowler}\\
      & T-Rittner &$-$ & $-$ &3.69& $-$ &\cite{Davis}\\
     & TEK$^a$ & $-$ & $-$ & 4.10& -1.64 &\cite{Davis}\\
     & D-shell\,(-1)$^b$&$-$ & $-$ & 4.27& -1.21 &\cite{Davis}\\
     & D-shell\,(q)$^c$ &$-$ & $-$ & 4.19& -0.02 & \cite{Davis}\\      
      \hline
 CaF & DF &  91.4 & 172.0 & 145.13 & -80.6 & This work\\
     & KRCISD &   92.5 & 151.0 & 131.50 & -58.5 & This work\\
      & T-Rittner &$-$ & $-$ &7.69& $-$ &\cite{Davis}\\
     & TEK$^a$ & $-$ & $-$ & 9.17& -5.77 &\cite{Davis}\\
          & D-shell\,(-1)$^b$&$-$ & $-$ &9.31&-5.20 &\cite{Davis}\\
          & D-shell\,(q)$^c$ &$-$ & $-$ &9.03& -4.03&\cite{Davis}\\  
 \hline
 SrF & DF\ &  137.0 & 248.5 & 211.33 & -111.5 & This work\\
      & KRCISD &  128.5 & 206.0 & 180.17 & -77.5 & This work\\
       & CCSD(T) &  126 & $-$ & $-$ & $-$ & \cite{Meyer}\\
      & CCSD(T) & 125 & $-$ & 170.05 & -66.35$^{\dag}$ & \cite{Kosicki}\\
       & T-Rittner &$-$ & $-$ &9.69 & $-$ &\cite{Davis}\\
     & TEK$^a$ & $-$ & $-$ & 11.69& -7.83 &\cite{Davis}\\
      & D-shell\,(-1)$^b$&$-$ & $-$ & 11.82& -7.21 &\cite{Davis}\\
      & D-shell\,(q)$^c$ &$-$ & $-$ &11.57& -6.35 &\cite{Davis}\\ 
 \hline
 BaF & DF &  214.5 & 413.5 & 339.33 & -222.5 & This work \\
      & KRCISD &  196.0 & 308.0 & 270.66 & -112.0 & This work\\
      & MRCISD & 182.590 & $-$ & $-$ & $-$ & \cite{Tohme}\\
      & T-Rittner &$-$ & $-$ &12.69& $-$&\cite{Davis}\\
      & TEK$^a$ & $-$ & $-$ &15.71& -11.41 &\cite{Davis}\\
      & D-shell\,(-1)$^b$&$-$ & $-$ &15.81& -10.72 &\cite{Davis}\\
      & D-shell\,(q)$^c$ &$-$ & $-$ &15.41& -9.50 &\cite{Davis}\\ 
 \hline
 RaF & DF &  195.0 & 351.0 & 299.0 & -156.0 & This work\\
     & KRCISD &  162.5 & 248.0 & 219.5& -85.5& This work\\
  \end{tabular}
\begin{flushleft}
$^a$T\"orring-Ernst-Kindt\\
$^b$Displaced shell model in which q$_+$ shell charges corresponding to valence shell electron, were set to -e.\\
$^c$Displaced shell model where q$_+$ shell charges were treated as parameters.\\
$^{\dag}$In Ref.~\cite{Kosicki}, the authors have reported positive value of $\gamma$ (or $\Delta\alpha$), however it comes out to be negative according to the definition given in eqn.~\ref{pol1}.
\end{flushleft}
\end{center}
\end{ruledtabular}
\end{table*}
\begin{table}[t]
\begin{ruledtabular}
\begin{center}
\caption{\label{table-VII} Results of DPs for the ground-state of AEMFs at equilibrium bond length with aug-QZ basis sets.}
\begin{tabular}{llllll}
%
Molecule & Method &  $\alpha$$_{\parallel}$ & $\alpha$$_{\bot}$ & $\bar{\alpha}$ & $\gamma$ \\
  \hline
BeF  & DF &   19.0 & 31.0 & 27.0&-12.0\\  
     & KRCISD & 20.5& 31.0 & 27.5& -10.5\\
MgF  & DF& 37.5 & 60.5 &52.83& -23.0\\  
     & KRCISD & 39.5& 61.0& 53.83& -21.5\\
CaF  & DF & 99.0 & 177.5 & 151.33&-78.5\\ 
     & KRCISD$^{*}$ & 100.0& 156.5& 137.67& -56.5\\
SrF  & DF & 139.5 & 249.5 &212.83&-110.0\\  
     & KRCISD$^{*}$ & 130.9& 207.0& 181.63 & -76.1 \\  
BaF  & DF & 221.0 & 422.0 & 355.0& -201.0\\  
     & KRCISD$^{*}$ & 203.9& 316.5& 278.97 &-112.6 \\
RaF  & DF & 202.0 & 358.5 &306.33 & -156.5\\ 
     & KRCISD$^{*}$ & 169.5& 255.5&226.83 & -86.0\\         
  \end{tabular}
\begin{flushleft}
$^{*}$ These values are calculated by adding the difference between the SCF energies calculated using the basis sets with and without augmentation, to the energies calculated at KRCISD level.\\
\end{flushleft}
\end{center}
\end{ruledtabular}
\end{table}  
\subsection{Atomic dipole polarizabilities ($\alpha_A$)}
Our results on $\alpha_A$'s for the alkaline-earth atoms, and fluorine together with the available results for these atoms in the literature are presented in Table~\ref{table-III}. The magnitude of $\alpha_A$ increases as we go from Be to Ba.  However, $\alpha_A$ decreases suddenly for Ra by about $8.2$\% when compared to Ba. This decrease in $\alpha_A$ is attributed to the fact that the value of $\left<r\right>$ for 7s of Ra is smaller than 6s of Ba~\cite{Chattopadhyay}. The correlation contributions to the atomic polarizabilities of alkaline-earth atoms at the level of CISD seem to be negative with reference to the DF results. Our results at the KRCISD/QZ level of correlation are in good agreement with the experimental data, well within their reported uncertainty limits~\cite{Peter, Ma, Miller, Miller1}. For Be and Ra, however, there are no experimental results available. Our results also compare quite well with the available theoretical calculations~\cite{Sahoo, Chattopadhyay, Porsev, Lim, Roos, Borschevsky}.

In addition, we have investigated the effect of diffuse functions on the results by repeating the calculations with singly augmented QZ basis sets. These additional functions do not seem to alter the value of $\alpha_A$ much, particularly at the DF level. The maximum change is only about $0.6$\% for Mg. However, at the KRCISD level, the maximum change obtained is in the result of $\alpha_A$ is $4.1$\% for Be.

The relativistic CCSD calculations have been reported in Ref.~\cite{Sahoo}, for alkaline-earth atoms along with He and Yb, and the contribution from the leading-order triples are quoted as error bars. Our KRCISD results compare quite well with those reported in Ref.~\cite{Sahoo}; the difference in the two being: $2.1$\% for Be, $0.6$\% for Mg, $1.6$\% for Ca, $2.9$\% for Sr and $0.5\%$ for Ba. Our results at KRCISD level also agree well, to within $3.2$\%, with the results reported in Ref.~\cite{Chattopadhyay} wherein they have used perturbed relativistic coupled\,-\,cluster (PRCC) method.

The polarizability calculations with the inclusion of core\,-\,core, and core\,-\,valence correlations using second\,-\,order many\,-\,body perturbation theory (MBPT2) and valence\,-\,valence correlations using the CI method has been reported in Ref.~\cite{Porsev}. Further, those authors have improved their results by obtaining the matrix elements using experimental lifetimes of the electronic states. Our \emph{ab initio} results reported in this work are in good agreement with their recommended values, with the maximum deviation being $2.4$\%.  

The DF results of $\alpha_A$'s for Ca through Ra, reported in our work, show excellent agreement with those given in Ref.~\cite{Lim} at the similar level. However, our fully relativistic KRCISD results vary from their recommended scalar relativistic results computed using the CCSD(T) method by $0.6$\% for Ca, $2.5$\% for Sr, $1.5$\% for Ba, and $0.5$\% for Ra.

The calculated value of polarizability for F with and without augmentation comes out to be $3.5$\,a.u. and $3$\,a.u., respectively, at the KRCISD level. The former value is in good agreement with the other reported calculations~\cite{Das, Medved, Nelin}. 

\subsection{Molecular permanent dipole moments}
The PDM values calculated using QZ quality basis sets in this work together with the available experimental, theoretical, and semi\,-\,empirical model-based results are tabulated in Table~\ref{table-IV} and our PDM results with aug-QZ basis sets are reported in Table~\ref{table-V}. Leaving aside BaF, our values of PDMs of all other AEMFs calculated at the KRCISD level show excellent agreement with the available results in the literature. To the best of our knowledge, the PDM of RaF molecule has not been reported earlier in the literature.

It has been observed that the correlation contributions to the PDMs of lighter molecules in this series are negative (-$0.13$\,D for BeF, and -$0.083$\,D for MgF), while for the heavier molecules, these contributions are comparatively large and positive ($0.287$\,D for CaF, $0.465$\,D for SrF, $0.584$\,D for BaF, and $0.576$\,D for RaF). At the level of KRCISD, the magnitude of PDM increases gradually from BeF to RaF, with BaF as an exception. Further, the effect of adding diffuse functions to QZ basis sets on PDMs of these molecules is not more than $1.2$\% at the DF level. However, at the KRCISD level there is a maximum of $2.4$\% increase in the value of PDM among the four lowest members of this series. Due to the lack of sufficient computational resources, we were unable to perform calculations with the augmented basis sets for BaF and RaF. In the discussion that follows, we have compared our results with the available experimental and other \emph{ab initio} results. 

Our calculated PDM results differ from the corresponding experimental results by $\sim3.6$\% for CaF~\cite{Childs}, $2.1$\% for SrF~\cite{Ernst}, and $14.6\%$ for BaF~\cite{Ernst1}. Further, it has to be noted that all available calculations including ours for BaF show a large deviation, ranging between $6.7$\% to $14.6$\%, from the experimental result reported in Ref.~\cite{Ernst1}.

Recently, Prasannaa~\emph{et al.}~\cite{Prasannaa} have performed CCSD calculations of PDMs by using only the linear terms in the CC wavefunction together with the cc\,-\,pVXZ (X\,=\,D,\,T,\,Q) basis sets for lighter elements (Be, Mg, Ca and F) and combination of Dyall- and Sapporo basis sets for Sr and Ba atom. In Table~\ref{table-IV}, we have quoted their results only at the QZ basis level for a fair comparison with our results and their results differ from ours by $0.016$\,D, $0.054$\,D, $0.021$\,D, and -$0.205$\,D for the first four members of the AEMF series. These differences lie almost within the error bars that they have reported ($\pm\,0.1$\,D for BeF, MgF, and CaF, while $\pm\,0.2$\,D for SrF). The PDM for BaF in our work deviates from their result by $0.694$\,D, which falls outside their estimated error bar of $0.2$\,D. However, they claim that their error estimate is completely uncertain due to numerical convergence issues and incompleteness of basis. 

The FFCCSD results for PDMs calculated by keeping all filled\,- and virtual orbitals as active have been reported in Ref.~\cite{Abe}. The basis sets used in their work are the same as that of Ref.~\cite{Prasannaa} discussed in the preceding paragraph. The difference between the two results is only $0.034$\,D for BeF, $0.006$\,D for MgF, $0.009$\,D for CaF, and $0.225$\,D for SrF. Furthermore, other recent calculations of PDM of SrF using the Z\,-\,vector method within CCSD~\cite{Sasmal}, and CCSD(T) approximation~\cite{Kosicki} differ from our KRCISD result by about $1.6$\% and $1.8$\%, respectively.

The absolute difference between our values of PDM at the DF level and those reported by Kobus~\emph{et al.}~\cite{Kobus} at the finite difference HF (FD\,-\,HF) level increases from BeF to SrF. Similar is the situation between our relativistic findings at KRCISD/QZ level and the results reported by Langhoff~\emph{et al.}~\cite{Langhoff} using CISD and coupled pair functional (CPF) method together with the extended Slater-type basis sets. The CPF results of ~\cite{Langhoff} seem to be more closer to our KRCISD results as compared to their CISD results. It has to be noted that in both works~\cite{Kobus} and \cite{Langhoff} the calculations have been performed in the non\,-\,relativistic framework. Therefore, the increase in difference between our results and those reported in the Refs.~\cite{Kobus, Langhoff} could be due to the increase in relativistic effects as one go from BeF to SrF.
 
\subsection{Molecular dipole polarizabilities ($\alpha$)}
The results for polarizability components of AEMFs calculated in this work using QZ basis sets are tabulated along with those of other calculations in Table~\ref{table-VI}. The magnitudes of both polarizability components ($\alpha_{\parallel}$ and $\alpha_{\bot}$) increase as one goes from BeF to BaF. However, it has been observed that the magnitudes of $\alpha_{\parallel}$ and $\alpha_{\bot}$ for RaF are smaller than those of BaF. The contribution of fluorine to the DP of these molecules is small when compared to the contribution of alkaline-earth atom. The decrease in DP of Ra when compared to Ba can be explained in terms of the structure of radium atom itself, as discussed in the case of atomic polarizability.

All molecules of group\,-\,II monofluorides exhibit negative polarizability anisotropy ($\gamma$), which is supported by \emph{ab initio} calculations in Ref.~\cite{Fowler} and by several model-based predictions in Ref.~\cite{Davis}. All available calculations including ours for polarizability components are higher than those predicted by Davis~\cite{Davis} using different models. 

The absolute value of correlation contribution to the average component of polarizability increases drastically as one traverses to higher members of AEMFs. Further, the correlation contributions for both $\bar{\alpha}$ and $\gamma$, as it can be seen from Table~\ref{table-VI}, are quite significant for the heavier molecules, \emph{viz.} CaF through RaF. The values of correlation contributions to the $\bar{\alpha}$ are: $13.63$\,a.u. for CaF, $31.16$\,a.u. for SrF, $68.67$\,a.u. for BaF, and $79.50$\,a.u. for RaF. These values are one order of magnitude larger than the correlation effects observed in MgF (=\,-\,$1.33$\,a.u.) and by two order larger than that observed for the case of BeF (=\,-\,$0.67$\,a.u.). Similarly, for $\gamma$, these contributions are $22.1$\,a.u. for CaF, $34$\,a.u. for SrF, $110.5$\,a.u. for BaF, and $70.5$\,a.u. for RaF, which are much larger than the value of $0.5$\,a.u. ($2.5$\,a.u.) for BeF (MgF) molecule.

The results of molecular polarizabilities using aug-QZ basis sets are tabulated in Table~\ref{table-VII}. We have observed that the augmentation of basis sets gives a positive contribution to the values of both $\alpha_{\parallel}$ and $\alpha_{\bot}$ components and is more effective for lower members of AEMFs rather than their heavier cousins at the DF level. Further, at the KRCI level, the augmented basis sets alter the value of $\alpha_{\parallel}$ by $7.9$\% ($5.3$\%) and $\alpha_{\bot}$ by $5.1$\% ($5.2$\%) for BeF (MgF) molecule. For other molecules, KRCISD calculations with the augmented basis sets are not performed due to the lack of sufficient computational resources. Nevertheless, we have estimated the effect of diffuse functions by adding the difference between the SCF energies calculated using the basis sets with and without augmentation, to the energies calculated at KRCISD level. With this energy correction, the estimated values of $\alpha_{\parallel}$ and $\alpha_{\bot}$ components increase, respectively, by $8.1$\% and $3.6$\% for CaF, $1.9$\% and $0.5$\% for SrF, $4$\% and $2.8$\% for BaF, and $4.3$\% and $3$\% for RaF. 

Fowler and Sadlej~\cite{Fowler} have employed the CASSCF method together with the contracted basis sets ($10$s\,$6$p\,$4$d) $\rightarrow$ [$5$s\,$3$p\,$2$d] for Be and F atom, and ($13$s\,$10$p\,$4$d) $\rightarrow$ [$7$s\,$5$p\,$2$d] for Mg, to compute the dipole polarizability of BeF and MgF molecules. A total of $9$ electrons were kept active in $4\sigma$ and $2\pi$ orbitals. We have also included the same number of electrons ($9$) in our correlation calculation of BeF and MgF. However, our virtual space ($80$ and $84$ molecular orbitals for BeF and MgF, respectively) is very large compared to that considered in Ref.~\cite{Fowler}. For BeF, our values of $\bar{\alpha}$ and $\gamma$ at KRCISD level differ from those reported in Ref.~\cite{Fowler} by $4.8$\% and -$0.5$\%, respectively, while for MgF, the difference between the two works is $4.1$\% for $\bar{\alpha}$ and $10$\% for $\gamma$. 

Meyer and Bohn~\cite{Meyer} have reported only the parallel component of polarizability for SrF using CCSD(T) method, available in MOLPRO package. They have used effective core potential (ECP28MDF) basis set for Sr and augmented correlation-consistent valence quadruple zeta (AVQZ) basis set for F atom. Our value of $\alpha_{\parallel}=128.5$\,a.u. at KRCISD level of correlation agrees well with the value of $126$\,a.u. reported in Ref.~\cite{Meyer} to within $2$\%. Quite recently, Kosicki~\emph{et al.}~\cite{Kosicki} have applied the  CCSD(T) method in conjunction with the core\,-\,valence aug\,-\,cc\,-\,pCVQZ basis set for F and ECP28MDF basis set for Sr atom to calculate the electric properties of SrF. The values of $\bar\alpha$ and $\gamma$ reported in Ref.~\cite{Kosicki} differ from ours by $5.6$\% and $14.4$\%, respectively. It must be mentioned here that we have included $17$ electrons to perform the correlation calculations in a virtual space of $132$ orbitals.

Tohme and Korek~\cite{Tohme} have computed the \textit{z}\,-\,component of polarizability for BaF molecule using MRCISD method. They have used ECP basis set ($5$s\,$5$p\,$4$d) for the Ba atom and contracted  basis set, ($6$s\,$3$p) $\rightarrow$ [$3$s\,$2$p] for F atom. Further, only five electrons are included in the correlation calculations. On the other hand, we have considered $17$ electrons as active in a virtual space of $134$ orbitals. The value of $\alpha_{\parallel}$ reported in Ref.~\cite{Tohme} is smaller by $6.8$\% from our result at the similar level of approximation.

The computed values of $\alpha_{\parallel}$ and $\alpha$$_{\bot}$ for CaF\,(RaF) are $92.5$\,($162.5$)\,a.u. and $151.0$\,($248.0$)\,a.u., respectively, by considering $15$\,($17$) electrons in the correlation calculation with $139$ ($133$) virtual orbitals. There is no calculation available in the literature, known to our knowledge, to compare the values of polarizability components of CaF and RaF, and perhaps we are contributing these results to the literature for the first time.

\section{\label{section-4}Summary}
In summary, we have applied finite\,-\,field method to carry out the PDMs and dipole polarizability calculations of alkaline-earth monofluorides and also, the atomic polarizability calculations for the atoms that form the diatomic molecules studied in this work, at the KRCISD level of correlation. These \emph{ab initio} calculations are performed using GAS technique with the uncontracted quadruple-zeta quality basis sets. Some of the results, particularly, the PDM of RaF and components of molecular polarizability for CaF and RaF are computed and reported in this work for the first time, to the best of our knowledge. The other findings for the atomic and molecular properties are in good agreement with the available results in the literature. Further, the effect of adding diffuse functions on the values of electric properties are examined by considering the singly augmented QZ basis sets. Except F for which the effect of diffuse functions to the $\alpha_A$ is $16.7$\%, the change in the values of $\alpha_A$'s for other atoms is not more than $4.1$\% at the KRCI level. The molecular PDMs results alter by a maximum of $2.4$\% among first four members of the series, using aug-QZ basis sets. However, the contribution of diffuse functions are quite significant for the components of molecular polarizability ranging between $1.9$\% to $7.9$\% for $\alpha_{\parallel}$ and $0.5$\% to $5.2$\% for $\alpha_{\bot}$.  We believe that our relativistic calculations performed with the large active space and optimized basis sets would be useful for the future theoretical studies and for the experimentalists working on the laser spectroscopy or the collision physics of these molecules.

\begin{center}
{\bf {ACKNOWLEDGMENTS}}
\end{center}
We thank Prof. M. Kajita of NICT, Japan and Dr. M. Abe of TMU, Japan, for helpful discussions. All calculation reported in this work were performed on the computing facility available in the Department of Physics at IIT Roorkee, India.\\

\begin{thebibliography}{99}
\bibitem{Yin} Y. Yin, Y. Xia, X. Li, X. Yang, S. Xu, and J. Yin, Applied Physics Express \textbf{8}, 092701 (2015).
\bibitem{Tarbutt} M. R. Tarbutt, and T. C. Steimle, Phys. Rev. A \textbf{92}, 053401 (2015).
\bibitem{Zhelyazkova} V. Zhelyazkova, A. Cournol, T. E. Wall, A. Matsushima, J. J. Hudson, E. A. Hinds, M. R. Tarbutt, and B. E. Sauer, Phys. Rev. A \textbf{89}, 053416 (2014).
\bibitem{Barry} J. F. Barry,  D. J. McCarron, E. B. Norrgard, M. H. Steinecker, and D. DeMille, Nature \textbf{512}, 286 (2014).
\bibitem{Barry1} J. F. Barry, E. S. Shuman, E. B. Norrgard, and D. DeMille, Phys. Rev. Lett. \textbf{108}, 103002 (2012).
\bibitem{Shuman} E. S. Shuman, J. F. Barry, and D. DeMille, Nature \textbf{467}, 820 (2010).
\bibitem{Truppe} S. Truppe, H. J. Williams, M. Hambach, L. Caldwell, N. J. Fitch, E. A. Hinds, B. E. Sauer, and M. R. Tarbutt Nat. Phys. \textbf{13}, 1173 (2017).
\bibitem{Bu} W. Bu, T. Chen, G. Lv, and B. Yan, Phys. Rev. A \textbf{95}, 032701 (2017).
\bibitem{Altunta} E. Altunta, J. Ammon, S. B. Cahn, and D. DeMille Phys. Rev. A \textbf{97}, 042101 (2018).	
\bibitem{Nayak} M. K. Nayak, and R. K. Chaudhuri, J. Phys. B: At. Mol. Opt. Phys. \textbf{39}, 1231 (2006).
\bibitem{Kudashov} A. D. Kudashov, A. N. Petrov, L. V. Skripnikov, N. S. Mosyagin, T. A. Isaev, R. Berger, and A. V. Titov Phys. Rev. A \textbf{90}, 052513 (2014).
\bibitem{Isaev} T. A. Isaev, S. Hoekstra, and R. Berger, Phys. Rev. A \textbf{82}, 052521 (2010).
\bibitem{Meyer} E. R. Meyer, and J. L. Bohn, Phys. Rev. A \textbf{83}, 032714 (2011).
\bibitem{Menotti} C. Menotti, M. Lewenstein, T. Lahaye, and T. Pfau, AIP Conf. Proc. \textbf{970}, 332 (2008).
\bibitem{DeMille} D. DeMille, Phys. Rev. Lett. \textbf{88}, 067901 (2002).
\bibitem{Meyer1} E. R. Meyer, J. L. Bohn, and M. P. Deskevich Phys. Rev. A \textbf{73}, 062108 (2006).
\bibitem{Trefzger} C. Trefzger, C. Menotti, B. Capogrosso-Sansone, and M. Lewenstein,  J. Phys. B: At. Mol. Opt. Phys. \textbf{44}, 193001 (2011).
\bibitem{Goral} K. G\'oral, L. Santos, and M. Lewenstein, Phys. Rev. Lett. \textbf{88}, 170406 (2002).
\bibitem{Friedrich} B. Friedrich, and D. Herschbach, Phys. Rev. Lett. \textbf{74}, 4623 (1995).
\bibitem{Friedrich1} B. Friedrich, and D. Herschbach, J. Phys. Chem. \textbf{99}, 15686 (1995).
\bibitem{Friedrich2} B. Friedrich, and D. Herschbach, J. Chem. Phys. \textbf{111}, 6157 (1999).
\bibitem{Friedrich3} B. Friedrich, and D. Herschbach, J. Phys. Chem. A \textbf{103}, 10280 (1999).
\bibitem{Deiglmayr} J. Deiglmayr, M. Aymar, R. Wester, M. Weidem\"uller, and O. Dulieu, J. Chem. Phys. \textbf{129}, 064309 (2008).
\bibitem{Tohme} S. N. Tohme, and M. Korek, J. Quant. Spectrosc. Radiat. Transfer \textbf{167}, 82 (2015).
\bibitem{Jardali} F. Jardali, M. Korek, and G. Younes, Can. J. Phys. \textbf{92}, 1223 (2014).
\bibitem{Kork} N. El-Kork,  N. A. el kher, F. Korjieh, J. A. Chtay, and M. Korek, Spectrochimica Acta Part A: Molecular and Biomolecular Spectroscopy \textbf{177}, 170 (2017).
\bibitem{Abe} M. Abe, V. S. Prasannaa, and B. P. Das, Phys. Rev. A \textbf{97} 032515 (2018).
\bibitem{Fowler} P. W. Fowler, and A. J. Sadlej, Molecular Physics \textbf{73}, 43 (1991).
\bibitem{Davis} S. L. Davis, J. Chem. Phys. \textbf{89}, 3 (1988).
\bibitem{Kosicki} M. B. Kosicki, D. Kedziera, and P. S. \.Zuchowski, J. Phys. Chem. A  \textbf{121}, 4152 (2017).
\bibitem{DIRAC}  DIRAC, a relativistic ab initio electronic structure program, Release DIRAC15 (2015), written by R. Bast, T. Saue, L. Visscher, and H. J. Aa. Jensen, with contributions from V. Bakken, K. G. Dyall, S. Dubillard, U. Ekstroem, E. Eliav, T. Enevoldsen, E. Fasshauer, T. Fleig, O. Fossgaard, A. S. P. Gomes, T. Helgaker, J. Henriksson, M. Ilias, Ch. R. Jacob, S. Knecht, S. Komorovsky, O. Kullie, J. K. Laerdahl, C. V. Larsen, Y. S. Lee, H. S. Nataraj, M. K. Nayak, P. Norman, G. Olejniczak, J. Olsen, Y. C. Park, J. K. Pedersen, M. Pernpointner, R. Di Remigio, K. Ruud, P. Salek, B. Schimmelpfennig, J. Sikkema, A. J. Thorvaldsen, J. Thyssen, J. van Stralen, S. Villaume, O. Visser, T. Winther, and S. Yamamoto (see http://www.diracprogram.org).
\bibitem{Dunning} T. H. Dunning, Jr., J. Chem. Phys. \textbf{90}, 1007 (1989).
\bibitem{Dyall_basis} K. G. Dyall, J. Phys. Chem. A \textbf{113}, 12638 (2009).
\bibitem{Chattopadhyay} S. Chattopadhyay, B. K. Mani, and D. Angom, Phys. Rev. A \textbf{89}, 022506 (2014).
\bibitem{Peter} P. Schwerdtfeger, Table of experimental and calculated static dipole polarizabilities for the electronic ground states of the neutral elements (in atomic units) 2015.
\bibitem{Ma} L. Ma, J. Indergaard, B. Zhang, I. Larkin, R. Moro, and W. A. de Heer, Phys. Rev. A \textbf{91}, 010501(R) (2015).
\bibitem{Miller} T. M. Miller, and B. Bederson, Measurement of the polarizability of calcium, Phys. Rev. A \textbf{14}, 1572 (1976).
\bibitem{Miller1} T. M. Miller, in CRC Handbook of Chemistry and Physics, Ed. D. R. Lide (CRC Press New York, 2002).
\bibitem {Sahoo} B. K. Sahoo, and B. P. Das, Phys. Rev. A \textbf{77}, 062516 (2008).
\bibitem{Porsev}S. G. Porsev, and A. Derevianko, JETP \textbf{102}, 195 (2006).
\bibitem{Lim} I. S. Lim, and P. Schwerdtfeger, Phys. Rev. A \textbf{70}, 062501 (2004).
\bibitem{Borschevsky} A. Borschevsky, V. Pershina, E. Eliav, and U. Kaldor, Phys. Rev. A \textbf{87}, 022502 (2013).
\bibitem{Roos} B. O. Roos, V. Veryazov, and Per-Olof Widmark, Theor. Chem. Acc.  \textbf{111} 345 (2004).
\bibitem{Das} A. K. Das, and A. J. Thakkar, J. Phys. B: At. Mol. Opt. Phys. \textbf{31}, 2215 (1998).
\bibitem{Medved} M. Medve$\check {d}$, P. W. Fowler, and J. M. Hutson, Molecular Physics, \textbf{98} 453 (2000).
\bibitem{Nelin} C. Nelin, B. O. Roos, and A. J. Sadlej, J. Chem. Phys. \textbf{77}, 3607 (1982).
\bibitem{Childs} W. J. Childs, L. S. Goodman, U. Nielsen, and V. Pfeufer, J. Chem. Phys., \textbf{80}, 2283 (1984).
\bibitem{Ernst} W. E. Ernst, J. K\"andler, S. Kindt, and T. T\"orring, Chem. Phys. Lett. \textbf{113}, 351 (1985). 
\bibitem{Ernst1} W. E. Ernst, J. Klindler, and T. T\"orring, J. Chem. Phys., \textbf{84}, 4769 (1986).
\bibitem{Prasannaa} V. S. Prasannaa, S. Sreerekha, M. Abe, V. M. Bannur, and B. P. Das, Phys. Rev. A \textbf{93}, 042504 (2016). 
\bibitem{Sasmal}S. Sasmal, H. Pathak, M. K. Nayak, Nayana Vaval, and S. Pal, Phys. Rev. A \textbf{91}, 030503(R) (2015).
\bibitem{Kobus} J. Kobus, D. Moncrieff, and S. Wilson, Phys. Rev. A \textbf{62}, 062503 (2000).
\bibitem{Langhoff} S. R. Langhoff, and C. W. Bauschlicher, Jr., J. Chem. Phys. \textbf{84}, 5025 (1986).
\bibitem{Buckingham} A. D. Buckingham, and R. M. Oleg\'ario, Chem. Phys. Lett. \textbf{212}, 253 (1993).
\bibitem{Torring} T. T\"orring, W. E. Ernst, and S. Kindt, J. Chem. Phys. \textbf{81}, 4614 (1984).
\bibitem{Rice} S. F. Rice, H. Martin, and R. W. Field, J. Chem. Phys. \textbf{82}, 5023 (1985).
\bibitem{Mestdagh} J. M. Mestdagh, and J. P. Visticot, Chemical Physics \textbf{155 }, 79 (1991).
\bibitem{Bundgen} P. B\"undgen, B. Engels, and S. D. Peyerimhoff, Chem. Phys. Lett. \textbf{176}, 407 (1991).
\bibitem{Allouche} A. R. Allouche, G. Wannous, and M. Aubert-Fr\'econ, Chem. Phys. \textbf{170}, 11 (1993).
\end{thebibliography}
\end{document}